\documentclass[12pt]{article}
\usepackage{amsmath, amssymb, amsfonts, color, latexsym, amscd}
\usepackage[dvips]{graphicx}
\usepackage{rotating}

\newcommand{\E}{\mathbb E}

\def\ind{ {{\rm 1}\hskip-2.2pt{\rm l}}}

\begin{document}

\title{An Extreme Value Theory approach for the early detection of time clusters with application to the surveillance of \textit{Salmonella}}

\author{\begin{normalsize}Armelle Guillou$^{(a)}$, Marie Kratz$^{(b)}$, Yann Le Strat$^{(c)}$\end{normalsize}}

\date{}
\maketitle

\begin{small}
\noindent $^{(a)}$ IRMA UMR 7501, Universit\'e de Strasbourg, France; Email: armelle.guillou@math.unistra.fr\\
$^{(b)}$ ESSEC Business School Paris \& MAP5 UMR 8145, Univ. Paris Descartes, France; Email: kratz@essec.fr\\
$^{(c)}$ Institut de Veille Sanitaire, D\'epartement des Maladies Infectieuses, Saint-Maurice, France; Email: y.lestrat@invs.sante.fr\newline
\end{small}

\begin{abstract}
\noindent We propose a method to generate a warning system for the early detection of time clusters applied to public health surveillance data. This new method relies on the evaluation of a return period associated to any new count of a particular infection reported to a surveillance system. 
The method is applied to Salmonella surveillance in France and  compared to the model developed by Farrington {\it et al.}
\end{abstract}

\noindent \footnotetext{
This work was partially supported by the ESSEC Research Center.\\~\\
\emph{2000 AMSC} 60G70 ; 62P10
}

\section{Introduction}
Since the pioneering work of Serfling (see \cite{Serfling63}), several statistical models have been proposed to detect time clusters from specific surveillance data. A time cluster is defined as a time interval in which the number of observed events is significantly higher than the expected number of events in a given geographic area. The term "event" is generic enough to include any event of interest such as a case of illness, an admission to an emergency department, a death or any other health event. 

\noindent The published models can be classified into three broad approaches: regression methods, time series methods and statistical process control as proposed by some recent reviews (see \cite{Sonesson03},\cite{Farrington04},\cite{LeStrat05}). In most cases they are based on two steps: (i) the calculation of an expected value of the event of interest for the current time unit (generally a week or a day); (ii) a statistical comparison between this expected value and the observed value. A statistical alarm is triggered if the observed value is significantly different from the expected value. 

\noindent The first step is based on the past counts, or more often on a sample of the past counts, that takes the seasonality pattern(s) into account. Thus, the current count is compared to counts that occurred in the past during the same time periods, e.g. the same week more or less three weeks for the last five years. Alternatively, sinusoidal seasonal components can be incorporated into regression models to deal with the seasonality and to easily take secular trends into account. More rarely, models try to reduce the influence of weeks coinciding with past outbreaks. One solution to avoid that such outbreaks reduce the sensitivity of the model is to associate low weights to these weeks (see e.g. \cite{Farrington96}).

\noindent The early prospective detection of time clusters represents a statistical challenge as the models must take the main feature of the data into account such as secular trends, seasonality, past outbreaks but are also faced with idiosyncrasies in reporting, such as delays, incomplete or inaccurate reporting or other artefacts of the surveillance systems. Reporting delays are particularly problematic for surveillance systems that are not based on electronic reporting. Concerning non-specific surveillance systems, the same difficulties are encountered, excepted for the reporting delays because these surveillance systems are mostly based on electronic reporting. 

\noindent The intentional release of anthrax in the USA in October 2001 emphasized the need to develop new early warning surveillance systems (see \cite{Goldenberg2002},\cite{Reis2003b}). These surveillance systems treat an increasing number of data provided from multiple sources of information (see \cite{CDC2004}). 
One logical consequence was to perform statistical analyses with a daily frequency.

\noindent Developing automated statistical prospective methods for the early detection of time clusters is thus essential. It is important for a public health surveillance agency to run several statistical methods concomitantly in order to compare the alarms generated by these methods. It is crucial to carry on the development of new methods because the combination of methods increases the sensitivity and the positive predictive value of the surveillance system.

\noindent It is the reason why we propose in this paper a new approach based on Extreme Value Theory (EVT) (see e.g. \cite{Embrechts01}) for the early detection of time clusters. To illustrate the performance of the method, we applied it to the detection of time clusters from weekly counts of \textit{Salmonella} isolates reported to the national surveillance system in France. 

\noindent Salmonellosis is a major cause of bacterial enteric illness in both humans and animals, with bacteria called \textit{Salmonella}. 
In France, \textit{Salmonella} is the first cause of laboratory confirmed bacterial gastroenteritis, of hospitalization and of death. In 2005, a study estimated that between 92 and 535 deaths attributable to non typhoidal \textit{Salmonella} occurred each year (see \cite{Vaillant2005}).

\noindent The paper is organized as follows. The surveillance system and the data are presented in Section 2. A description of our method to check if each new observation corresponds to an unusual/extremal event is given in Section 3. Applications to counts of \textit{Salmonella} as well as a comparison to the Farrington method (see \cite{Farrington96}) are developed in Section 4. A discussion follows in the last section.

\section{Data}

\begin{figure}[ht]
\begin{center}
\scalebox{0.75}[0.75]{\includegraphics{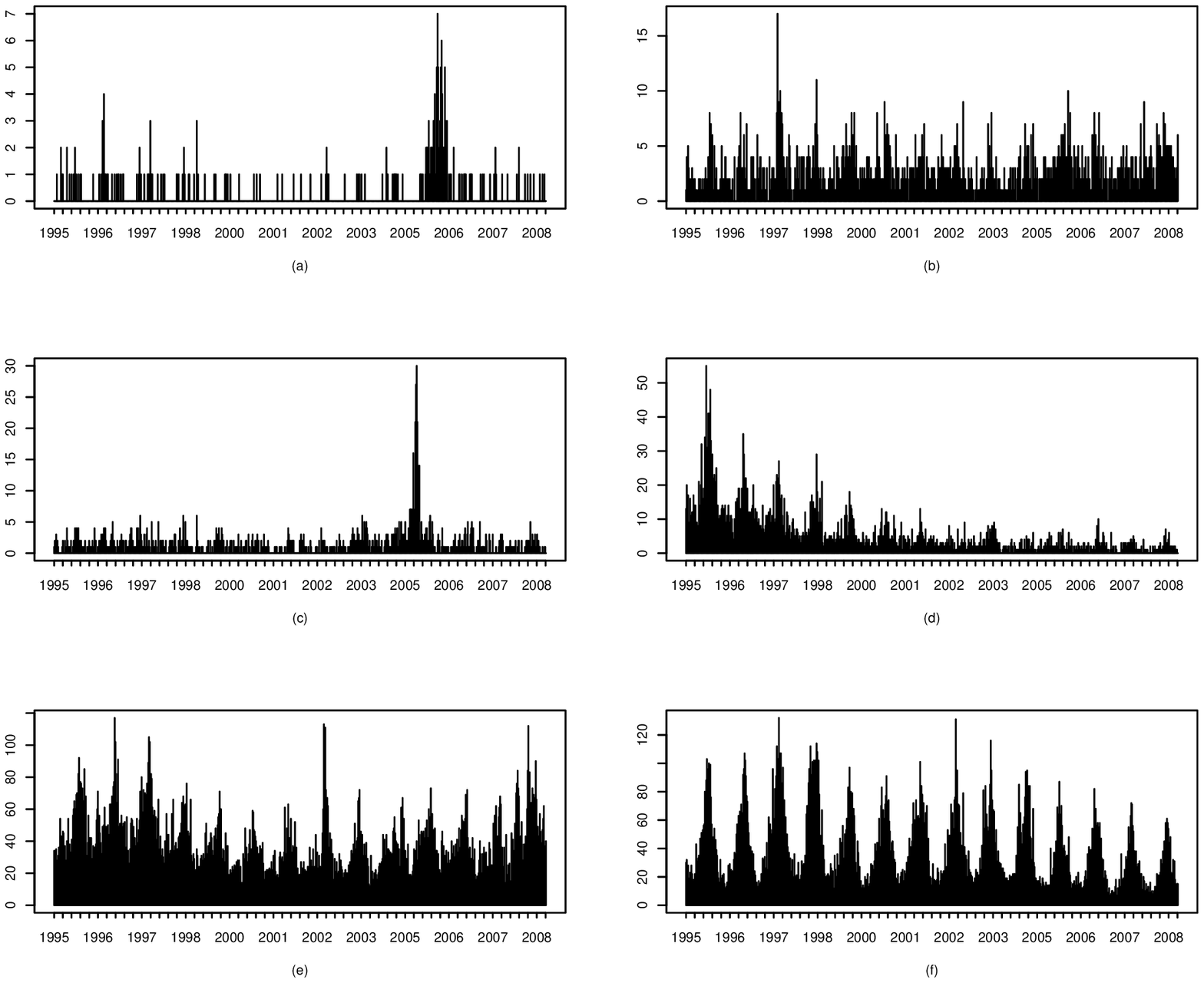}}
\caption{Weekly counts of isolates reported to the National Reference Center for \textit{Salmonella} in France, January 1, 1995  to December 31, 2008: (a) \textit{Salmonella} Manhattan; (b) \textit{Salmonella} Derby; (c) \textit{Salmonella} Agona; (d) \textit{Salmonella} Virchow; (e) \textit{Salmonella} Typhimurium; (f) \textit{Salmonella} Enteritidis.}\label{fig:Figure1}
\end{center}
\end{figure}

The National Reference Center for \textit{Salmonella} contributes to the surveillance of salmonellosis by performing serotyping of about 10000 clinical isolates each year. \textit{Salmonella} surveillance is based on a network of 1500 medical laboratories that voluntarily send their isolates. \textit{Salmonella} enterica serotypes Thyphimurium and Enteritidis represent 70 \% of all \textit{Salmonella} isolates in humans among many hundreds of serotypes; that is why we consider in this paper mainly these two serotypes. For illustrative purpose, four other less frequent serotypes (Manhattan, Derby, Agona and Virchow) might also be considered; Figure \ref{fig:Figure1} shows the weekly number of isolates for these six serotypes from January 1, 1995 to December 31, 2008. It highlights the great variability in terms of seasonality, secular trend and weekly number of reported isolates and frequencies of unusual events. \newline

\noindent Let $\{Y_t; t \in \mathbb N\}$ be the time series corresponding to the number of isolates at time point $t$ for a given serotype. As mentioned by many authors (see e.g. \cite{Nobre2001}, \cite{Goldenberg2002}), seasonal effects may have a strong impact to generate a statistical alarm. A usual way to prepare dataset is to select counts from comparable periods in past years, as described in the literature (see \cite{Stroup89}, \cite{Farrington96}). The dataset is restricted to the counts that occurred during the times within these comparable periods. For instance, if the current time is $t$ of year $y$, then only the counts for the $n=b(2w+1)$ times from $t-w$ to $t+w$ of years from $y-b$ to $y-1$ ($b>1$, $w>1$) are used. \\
From now on, $(X_t)$ will denote the resulting times series that will be considered in our study. As an illustration Figure 2 represents the restricted dataset for \textit{Salmonella} Typhimurium, for a given current week.

\begin{figure}[ht]
\begin{center}
\scalebox{0.5}[0.5]{\includegraphics[angle=-90]{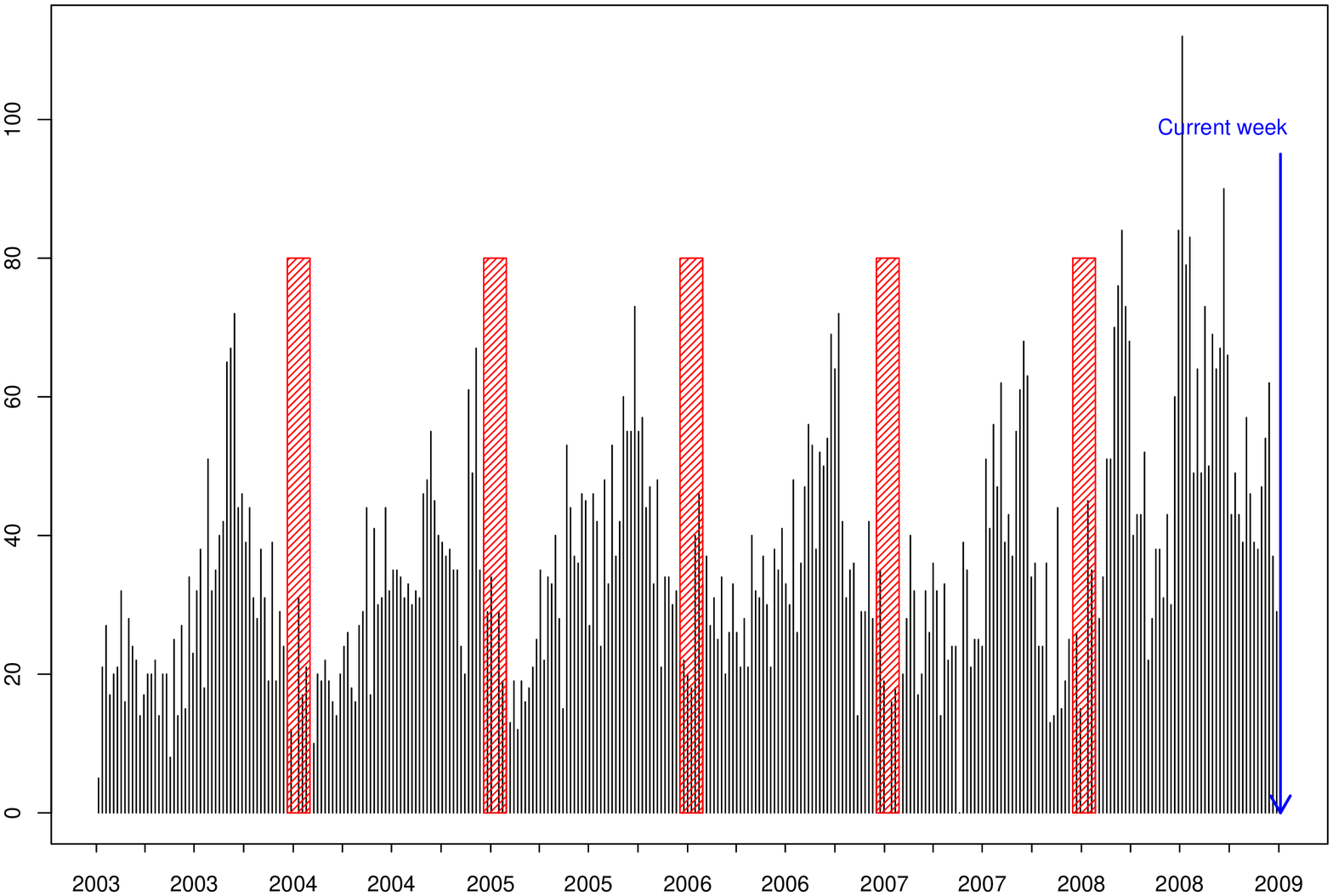}}
\caption{Weekly counts of isolates for \textit{Salmonella} Typhimurium. The current week, represented by the (blue) arrow, is the last week of December 2008. The counts 
that occurred in comparable periods ($\pm$ 3 weeks) in the five previous years, and used to generate or not an alarm, are represented by the (red) striped bands.}\label{fig:Figure2}
\end{center}
\end{figure}

\section{An EVT approach}

Suppose we have at our disposal $n$ successive observations that we consider as realizations of a sample $(X_i)$ of independent and identically distributed (i.i.d.) non-negative random variables defined on a probability space $(\Omega,\cal A, \mathbb P)$, from a distribution function $F$.\\
Recall that a return level $z_T$ associated with a given return period $T$ corresponds to the level  expected to be exceeded on average once every $T$ time units, 
{\it i.e.} such that 
$$
\E \left( \sum_{i=1}^{T} \ind_{\{X_i > z_T\}} \right ) =1
$$  
where $ {\ind}_{\{A\}}$ represents the indicator function that is  equal to $1$ if A is true and to $0$ otherwise.
The last equality can be rewritten as $1-F(z_T) = 1/T$.
Hence, the return level $z_T$ corresponds simply to a $p_T-$quantile  with $p_T=1-1/T$, $\displaystyle z_T=F^{\leftarrow}(1-1/T)$, $F^{\leftarrow}$ denoting the 
generalized inverse function of $F$.  \\
The idea of the method is to associate with each observation $x_{s}$ a return period $T_s$ defined theoretically as $(1-F(x_s))^{-1}$ to be able to determine 
the return period $T_t$ associated to each new observation $x_t$ at time $t$, then to look backwards (and not forwards as in the standard way) in the interval $(t-T_t; t)$ for the existence of an observation 
that would exceed $x_t$; if it exists, we generate a warning time at time $t$ since on average we were not expecting a second exceedance on $(t-T_t; t]$.\\
Notice that in our discrete framework it will not be possible to estimate explicitly the return levels; instead estimated bounds proposed in \cite{Guillou09} will be considered.\\
Therefore, after a preliminary analysis of the
data and definition of our sample, we will compute the estimated bounds
of the return levels in order to obtain a graph of the return periods
and levels. Then, we will allocate a return period to any new
observation $x_t$ to test if $t$ corresponds to a warning time according
to our definition.

\subsection{Bounds for the return level}

Looking at extremal events leads us to the crucial problem of high quantile estimation. Such a purpose has been extensively studied in the literature (see e.g. \cite{Embrechts01}), 
and the classical approach, in the i.i.d. setting, consists to use the Extreme Value Theory assuming that exceedances above a high threshold  approximately follow a Generalized Pareto distribution (this result is known as the Peak-Over-Threshold (POT) method). However, this theory is only valid  in the case where the underlying distribution function $F$ is continuous. 
This is not the case in the epidemiology context. Therefore, we propose to use instead upper and lower bounds for the return level $z_t$ and estimate them, 
following the method developed by Guillou {\it et al.} (see \cite{Guillou09}); this method has several advantages: the upper and lower bounds can be computed for any value of $t$ 
(in particular it holds for large values), it does work for both small and large samples, and for $F$ continuous or discrete. So this approach is 
well-adapted to our context, when assuming the random variables associated to the observations i.i.d.
\vskip1ex
\noindent Let us recall the expression of those upper and lower bounds, given respectively by
\begin{equation}\label{eq: upper bound 0}
\qquad \inf \Bigr\{  b_t(u,v): 
        \mbox{$u \geq 0 $, $v \geq 0$  non-decreasing functions} \Bigr\}\quad
\end{equation}
where $\displaystyle
b_t(u,v):=u^{\leftarrow}\left(\frac{t  \theta (u,v) } { v(1-1/t)} \right)\mbox{, with } \theta (u,v):= \E [u(X)v(F(X))]$,\\
and
\begin{equation}\label{eq: lower bound 0}
\sup \Bigr\{ \ell_t(u,w,q):\mbox{$u \geq 0$ non-decreasing function,$w\geq 0$ non-increasing function,$q>1$}\Bigr\}~
\end{equation}
where
$$
\ell_t(u,w,q) := u^{\leftarrow}\left(\frac{ \theta ^*(u,w)-  t^{-1+1/q}  \Bigr( \theta ^*(u^q,w^q)   \Bigr)^{1/q} } { w( 1/t) (1-1/t)} \right);~\theta ^*(u,w):= \E\left[ u(X)w(1-F(X))\right].$$

\noindent Estimators of those two bounds (\ref{eq: upper bound 0}) and  (\ref{eq: lower bound 0}) follow when considering natural estimators of $\theta (u,v)$ and $\theta ^*(u,w)$, namely
\begin{equation}\label{eq:estim-bds}
\widehat{b}_t(u,v) = u^{\leftarrow}\left[ \frac{t  \widehat{\theta}_n (u,v) } { v(1-1/t)} \right]
~\mbox{and}~
\widehat{\ell}_t(u,w,q) = u^{\leftarrow}\left[ \frac{ \widehat{\theta}_n ^*(u,w)-  t^{-1+1/q}  \left( \widehat{\theta}_n ^*(u^q,w^q)   \right)^{1/q} } { w( 1/t) (1-1/t)} \right], ~
\end{equation}
where, if $(X_{1,n}\le ...\le X_{n,n})$ denote the order statistics from a given sample,
$$
\widehat\theta _n(u,v) = {1 \over n} \sum_{i=1}^n u(X_{i,n})v\Bigr({i\over n}\Bigr) \quad\mbox{and}\quad
\widehat\theta ^*_n(u,w) = {1 \over n} \sum_{i=1}^{n} u(X_{i,n})w\Bigr(1-{i\over n}\Bigr).
$$
Under some conditions on $u$, $v$ and $w$, asymptotic distributions
are obtained for the bounds estimators, as well as asymptotic confidence intervals when using the delta method (see Section 3 in \cite{Guillou09}). \\
For instance, concerning the upper bound, we have the following asymptotic confidence interval:
\begin{equation}\label{eq: ci}
b_t(u,v) \in \Biggr[\widehat b_t(u,v) \pm {q_{1-\alpha/2} \, t \, \widehat \sigma \over \sqrt n \, v(1-{1\over t})} (u^{\leftarrow})'\Bigr({t\widehat \theta_n(u,v) \over v(1-{1\over t})}\Bigr)\Biggr]
\end{equation}
where $q_{1-\alpha/2}$ denotes the quantile of order $1-\alpha/2$ of the standard normal distribution and $\widehat \sigma$ the empirical version of $\sigma$ defined for $U$ uniformly distributed on $(0, 1)$ as \\ 
$\displaystyle 
\sigma^2 = \mathbb E\Biggr(-v(U)u(F^{\leftarrow}(U))+\theta (u,v) - \int_0^1\Bigr(\ind_{\{U\leq t\}}-t\Bigr)v^{\prime}(t)u(F^{\leftarrow}(t))dt\Biggr)^2$.\\
A similar confidence interval can be obtained for the lower bound.
\vskip2ex
\vskip1ex
\noindent
Finally, since it is impossible to optimize in (\ref{eq: upper bound 0}) and  (\ref{eq: lower bound 0}) under the whole family of non-negative and non-decreasing 
functions $u$ and $v$ and non-negative and non-increasing functions $w$, we reduce the problem by choosing the sub-class of power functions since it seems 
adapted to our study, giving reasonable results (even if not optimal).\\
We consider the functions defined by $u(x) =x^{\alpha}$, $v(x)=x^{\beta}$, $w(x)=x^{-\nu}$, with $\alpha,\beta,\nu$ positive real numbers and set $q=2$ 
(changing the value of this last parameter does not affect significantly 
the final result).
Then we solve numerically, for $\varepsilon$ close enough to 0, the following optimization problem
\begin{eqnarray}\label{minimization}
(\widehat{\alpha}_t,\widehat{\beta}_t) &=& \mbox{argmin} \; \Bigr\{\widehat{b}_t(x^\alpha,x^\beta) : \alpha\in[\varepsilon,\alpha_0], \beta\in[\varepsilon, \beta_0] \Bigr\} \quad\mbox{ and }\nonumber\\
\qquad (\widehat{\alpha}^{\, *}_t,\widehat{\nu}^{\, *}_t) &=& \mbox{argmax} \; \Bigr\{ \widehat{\ell}_t(x^\alpha,x^{-\nu},2)  : \alpha\in[\varepsilon, \alpha_0], \nu\in[\varepsilon,\nu_0] \Bigr\}, ~
\end{eqnarray}
in order to obtain the estimated upper and lower bounds equal, respectively, to 
\begin{equation}\label{eq: optimization}
\widehat b_t=\widehat{b}_t(x^{\widehat{\alpha}_t}, x^{{\widehat \beta}_t})\quad
\mbox{  and }\quad
\widehat{\ell}_t=\widehat{\ell}_t (x^{\widehat{\alpha}^{\, *}_t}, x^{-\widehat{\nu}^{\, *}_t},2). 
\end{equation}
As already said, the choice for $u,v,w$ and $q$ does not necessarily correspond to the optimal bounds but covers a wide enough range of bounds that provides satisfying results when working on various epidemiology datasets, as we are going to see in Section 4.

\subsection{Determination of an alarm time}

Let us present our method to define an alarm system.\\ 
It will consist in three main steps.\\ 
Note that using bounds for a return level $z_t$ will imply that the return period defined theoretically by $(1-F(z_t))^{-1}$ cannot be explicitly estimated and we have
\begin{equation}\label{comparisonTimes}
 T_{\ell}\le (1-F(z_t))^{-1}\le T_b
\end{equation}
where  $T_{\ell}$ and $T_{b}$ denote the return periods of the bounds $\ell_{t}(u,w,q)$ and $b_{t}(u,v)$ respectively.\\
\newline
\textbf{Step 1}: We draw the plot of the return period on the $x$-axis and the corresponding estimate of the upper bound of the return level (instead of the return level itself): $\left(t,\widehat{b}_{t}\right)$.\\
\newline
\textbf{Step 2}: We allocate to each observation $x_{t_i}$, $i\ge 1$, a return time $T_i$ using the previous plot. Namely, $x_{t_i}$ corresponds to a value $\widehat{b}_{T_i}$ of the $y$-axis of the plot from which we deduce the associated return level $T_i$. Reading an observation as an upper bound of a return level means that $T_i$ is in fact the lower bound of the theoretical return period $(1-F(x_{t_i}))^{-1}$ that should be associated to the observation $x_{t_i}$, because of (\ref{comparisonTimes}).\\
We justify our choice as follows. Considering $\widehat{\ell}_{T_i}$ instead of $\widehat{b}_{T_i}$ in the above method would have led to overestimate the return period associated to the observation $x_{t_i}$, which could be a problem in the context of alarms (it is better to have more alarms than less), except if the plots $\left(t,\widehat{\ell}_{t}\right)$ and $\left(t,\widehat{b}_{t}\right)$ were close enough, but it is generally not the case for our datasets where $\widehat{\ell}_{t}$ appeared approximately as a constant function of time (see \cite{Anis08}).\\
\newline
\textbf{Step 3}: We use the fact that if $(X_j)$ are i.i.d. random variables, then we have for any time interval $I\left(T\right)$ with length $T$
\begin{equation}\label{defreturn}
\mathbb E \left(\sum^{T}_{i=1} \ind_{\{X_{i}> z_T\}}\right)=1\Leftrightarrow \mathbb E \left(\sum_{i\in I\left(T\right)} \ind_{\{X_{i}> z_T\}}\right)=1.
\end{equation}
This remark is important since we want to define for each new observation a warning time and not a predicted return period, which means to look backward in time. \\
Hence for each new observation $x_{t_i}$, to which a return time $T_i$ has been associated (via Step 2), we will look in the interval $(t_i-T_i; t_i)$ to see if there exists an observation exceeding $x_{t_i}$; if it does, we ring an alarm at this time $t_i$, since as already said, on average, we do not expect a second exceedance on $(t_i-T_i; t_i]$.
Notice that we chose here to consider an inequality in the indicator set, and not a strict one as in (\ref{defreturn}), the level corresponding now to an observation.\\
To finish this section, let us sumarize our method.
\begin{itemize}
\item Using the plot $\left(t,\widehat{b}_{t}\right)$, associate a time $T_i$ to each observation $x_{t_i}$ ($i\ge 1$); 
\item for each new observation $x_{t_i}$, consider $I\left(T_i\right)=(t_i-T_i,t_i]$;
\item look for the existence of an observation $x_t \ge \widehat{b}_{T_i}$, for $t\in (t_i-T_i,t_i)$;\\ if there exists at least such an observation, {\it i.e.}\\ if $\displaystyle \mathbb E \left(\sum_{j\in I\left(T_i\right)} \ind_{\{ X_{j}\geq ~\widehat{b}_{T_i}\}}\right)>1$,
then generate a warning time at $t_i$.\\~
\end{itemize}

\noindent Now to illustrate our method, let us consider the example of the maximum number of \textit{Salmonella} Virchow isolates. In Figure 3, the $x$-axis corresponds to 
the values of $t$ from $2$ to $100$ weeks and the $y$-axis to $\widehat{b}_{t}$; the two dashed lines indicate the 95$\%$ confidence interval bounds of $b_t$. \\

\noindent The return level/return period graphs for the six serotypes presented in Section 2 are represented in Figure 4.
\begin{figure}[ht]
\begin{center}
\scalebox{0.65}[0.65]{\includegraphics{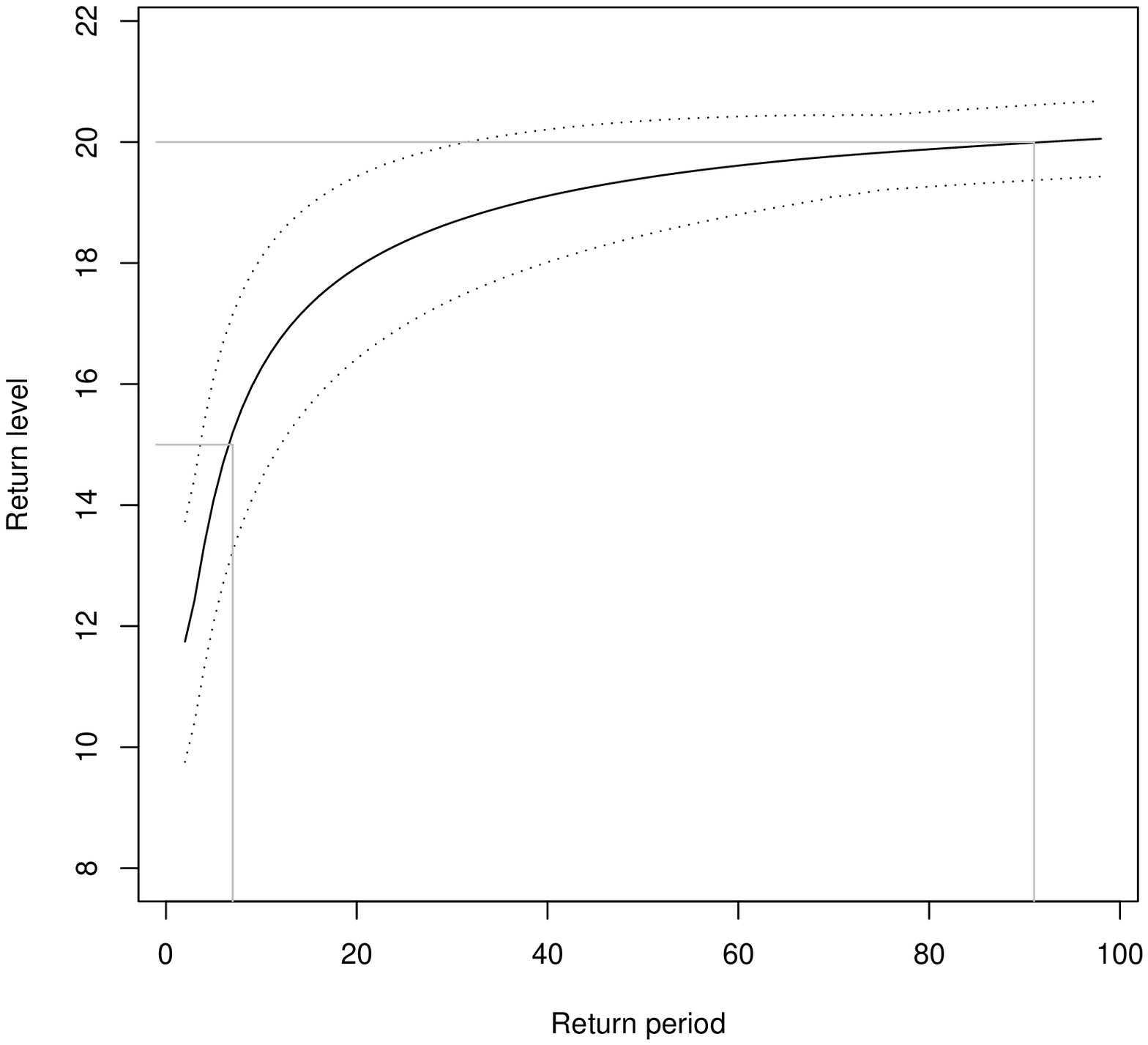}}
\caption{The return level/return period graph for \textit{Salmonella} Virchow. The black curve represents the upper bound of the return level. Dashed curves represent the 95$\%$ confidence interval of this upper bound. To the observation $y=20$ (respectively $y=15$) does correspond on the $x$-axis $\widehat{b}_{91}$ 
(respectively $\widehat{b}_{7}$) from which we deduce the return period equals to $T=91$ (respectively $T=7$) weeks.}\label{fig:Figure3}
\end{center}
\end{figure}
%
%
\begin{figure}[ht]
\begin{center}
\scalebox{0.7}[0.7]{\includegraphics{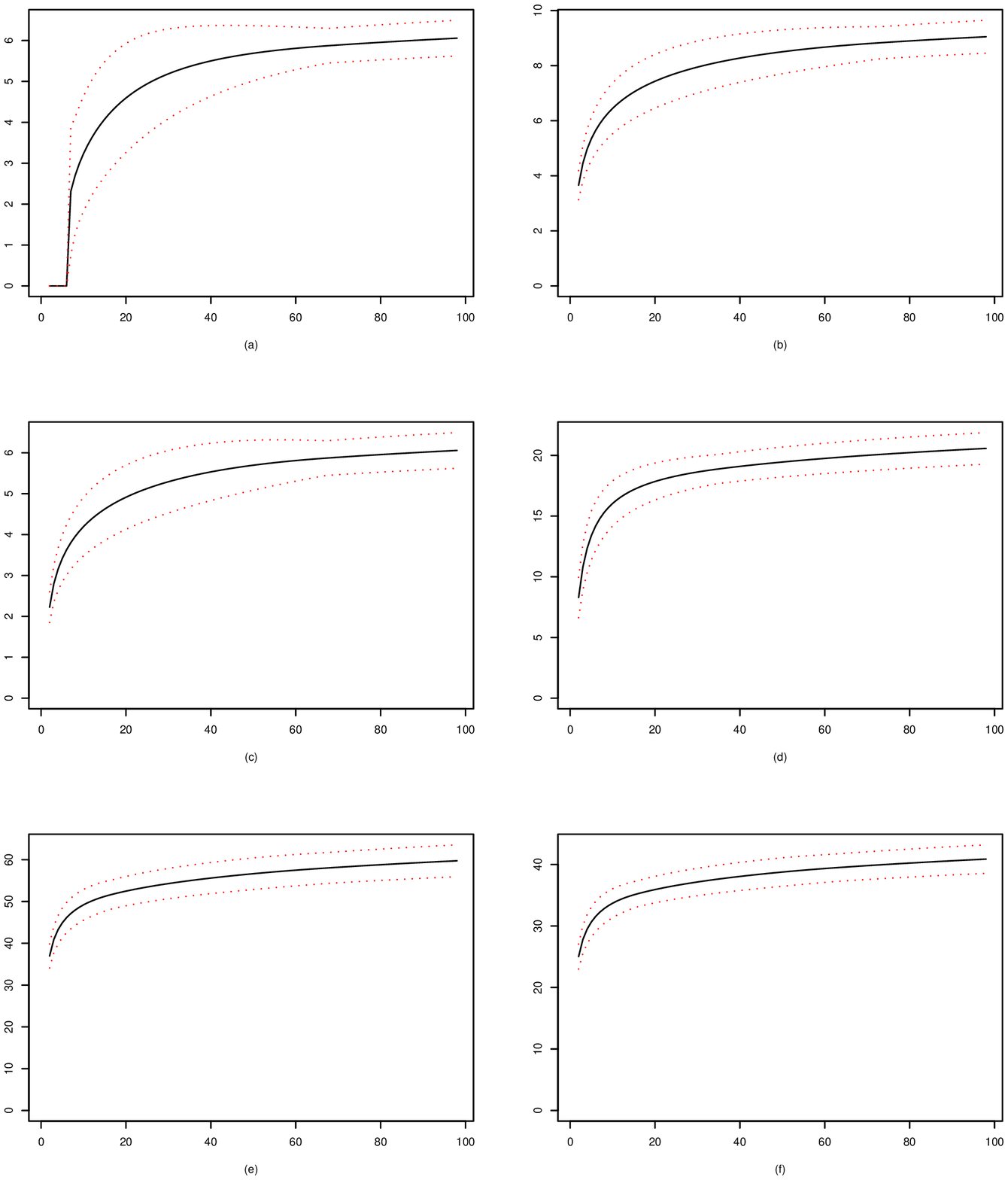}}
\caption{The return level/return period graphs calculated for the last week in 2008: (a) \textit{Salmonella} Manhattan; (b) \textit{Salmonella} Derby; (c) \textit{Salmonella} Agona; (d) \textit{Salmonella} Virchow; (e) \textit{Salmonella} Typhimurium; (f) \textit{Salmonella} Enteritidis.}\label{fig:Figure4}
\end{center}
\end{figure}

\section{Applications}

For each week from January 2000 to December 2008, the EVT method was applied to the six time series presented in Section 2. For any week $t$, counts of weeks 
from $t-3$ to $t+3$ of years from $y-5$ to $y-1$ were used. Moreover, in order to reduce the probability that an alarm could be triggered for few sporadic cases, a standard rule has been adopted to keep an alarm at week $t$ if at least 5 cases were observed during the 4 last weeks preceding $t$. This rule has already been applied at the Communicable 
Disease Surveillance Center (CDSC) using the method developed by Farrington {\it et al.} (see \cite{Farrington96}).

\noindent Several statistical methods for the  prospective change-point detection in time series of counts have been already applied to \textit{Salmonella} infections by various authors (see \cite{Watier91},\cite{Farrington96},\cite{Hutwagner97}). We chose to compare the alarms generated by the EVT method with the ones generated by the Farrington method for the six serotypes and for each week since 2005, as the National Reference Center for \textit{Salmonella} applies the Farrington method. 

\noindent Comparisons between the two methods are shown in Table 1. The concordance is equal to 93.7\% (Typhimurium), 95.9\% (Derby), 96.8\% (Agona), 97.4\% (Virchow), 98.7\% (Enteritidis) and 99.1\% (Manhattan). When comparing the results outside of the diagonal with the real past epidemies, it appears that our method seems to fit better the reality than Farrington's one, providing less alarms when historically there was indeed no epidemy, and more alarms when there was. It makes our method quite promising. 

\noindent Nevertheless 
this comparison cannot replace a more standard procedure that systematically compare for each week the statistical alarms and the alerts identified by an epidemiologist. In this case, the epidemiologist judgement is considered as the gold standard (even if the epidemiologist is not infallible).

\noindent Figures 5 and 6 represent both the alarms and the weekly counts over time for the serotypes Manhattan and Agona. Each triangle represents a statistical alarm. Triangles on the first line represent the alarms generated by the EVT method, whereas the ones generated by the Farrington method are given  on the second line.

\begin{table}[t]
\begin{center}
\begin{tabular}{rrrrrrrrrrr}
\hline
\multicolumn{5}{c}{Manhattan}                                                                                            & & \multicolumn{5}{c}{Derby} \\
\hline
                          &                  & \multicolumn{3}{c}{EVT}                                                   & &            & & \multicolumn{3}{c}{EVT}             \\
\cline{2-5}
\cline{8-11}
                          &                       & \multicolumn{1}{c}{-} & \multicolumn{1}{c}{+} & \multicolumn{1}{c}{Total} & & 
                          &                       & \multicolumn{1}{c}{-} & \multicolumn{1}{c}{+} & \multicolumn{1}{c}{Total} \\
                          & \multicolumn{1}{c}{-}      & 440 &   3 & 443 & &                           & \multicolumn{1}{c}{-}     & 441 &  12 & 453 \\
\multicolumn{ 1}{c}{Farrington} & \multicolumn{1}{c}{+}      &   1 &  19 &  20 & & \multicolumn{ 1}{c}{Farrington} & \multicolumn{1}{c}{+}     &   7 &   3 &  10 \\
                          &  \multicolumn{1}{c}{Total} & 441 &  22 & 463 & &                           & \multicolumn{1}{c}{Total} & 448 &  15 & 463 \\
                          &                            &     &     &     & &                           &                           &     &     &     \\

\hline
\multicolumn{5}{c}{Agona}                                                                                                & & \multicolumn{5}{c}{Virchow} \\
\hline
                          &                  & \multicolumn{3}{c}{EVT}                                                   & &            & & \multicolumn{3}{c}{EVT}             \\
\cline{2-5}
\cline{8-11}
                          &                       & \multicolumn{1}{c}{-} & \multicolumn{1}{c}{+} & \multicolumn{1}{c}{Total} & & 
                          &                       & \multicolumn{1}{c}{-} & \multicolumn{1}{c}{+} & \multicolumn{1}{c}{Total} \\
                          & \multicolumn{1}{c}{-}      & 427 &  13 & 440 & &                           & \multicolumn{1}{c}{-}     & 451 &   1 & 452 \\
\multicolumn{ 1}{c}{Farrington} & \multicolumn{1}{c}{+}      &   2 &  21 &  23 & & \multicolumn{ 1}{c}{Farrington} & \multicolumn{1}{c}{+}     &  11 &   0 &  11 \\
                          &  \multicolumn{1}{c}{Total} & 429 &  34 & 463 & &                           & \multicolumn{1}{c}{Total} & 462 &   1 & 463 \\
                          &                            &     &     &     & &                           &                           &     &     &     \\

\hline
\multicolumn{5}{c}{Typhimurium}                                                                                          & & \multicolumn{5}{c}{Enteritidis} \\
\hline
                          &                  & \multicolumn{3}{c}{EVT}                                                   & &            & & \multicolumn{3}{c}{EVT}             \\
\cline{2-5}
\cline{8-11}
                          &                       & \multicolumn{1}{c}{-} & \multicolumn{1}{c}{+} & \multicolumn{1}{c}{Total} & & 
                          &                       & \multicolumn{1}{c}{-} & \multicolumn{1}{c}{+} & \multicolumn{1}{c}{Total} \\
                          & \multicolumn{1}{c}{-}      & 418 &   1 & 419 & &                           & \multicolumn{1}{c}{-}     & 455 &   0 & 455 \\
\multicolumn{ 1}{c}{Farrington} & \multicolumn{1}{c}{+}      &  28 &  16 &  44 & & \multicolumn{ 1}{c}{Farrington} & \multicolumn{1}{c}{+}     &   6 &   2 &   8 \\
                          &  \multicolumn{1}{c}{Total} & 446 &  17 & 463 & &                           & \multicolumn{1}{c}{Total} & 461 &   2 & 463 \\
                          &                            &     &     &     & &                           &                           &     &     &     \\
\end{tabular} 
\end{center} 
\begin{center}
\caption{Two-way tables of frequency counts of non-alarms (-) and alarms (+) from the Farrington method and the EVT method, for each serotype.}
\end{center}
\end{table}

\noindent In Figure 5, the alarms generated by the two methods occurred in the same period that corresponds to a documented outbreak, delimited by the dashed lines, for 
the serotype Manhattan (see \cite{Noel2006}). From August 2005 to February 2006, a community-wide outbreak of \textit{Salmonella} 
Manhattan infections occurred in France. The investigation incriminated pork products from a slaughterhouse as being the most likely source of this outbreak. 
There was a concordance between the temporal (October-December 2005) and the geographical (south-eastern France) occurrence of the majority of cases and the 
distribution of products from the slaughterhouse.

\begin{figure}[ht]
\begin{center}
\scalebox{0.5}[0.5]{\includegraphics[angle=-90]{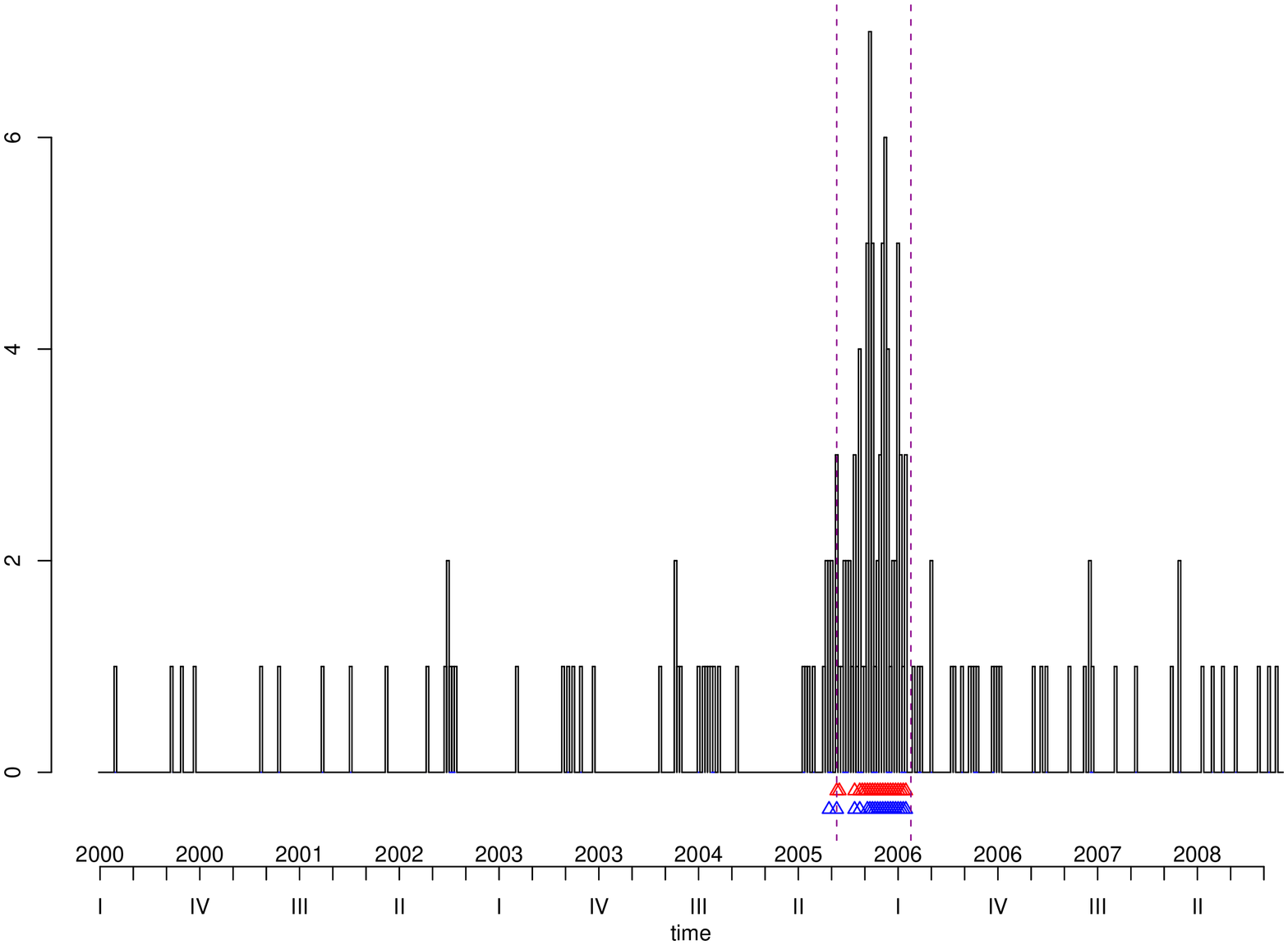}}
\caption{\textit{Salmonella} Manhattan: Weekly counts from January 1, 2000 to December 31, 2008. Roman numerals refer to the quarters of the years. 
Alarms generated by the EVT method are represented by triangles on the first line. Alarms generated by the Farrington method are represented by triangles 
on the second line. The documented outbreak is delimited by the two dashed lines.}\label{fig:Figure5}
\end{center}
\end{figure}

\begin{figure}[ht]
\begin{center}
\scalebox{0.5}[0.5]{\includegraphics[angle=-90]{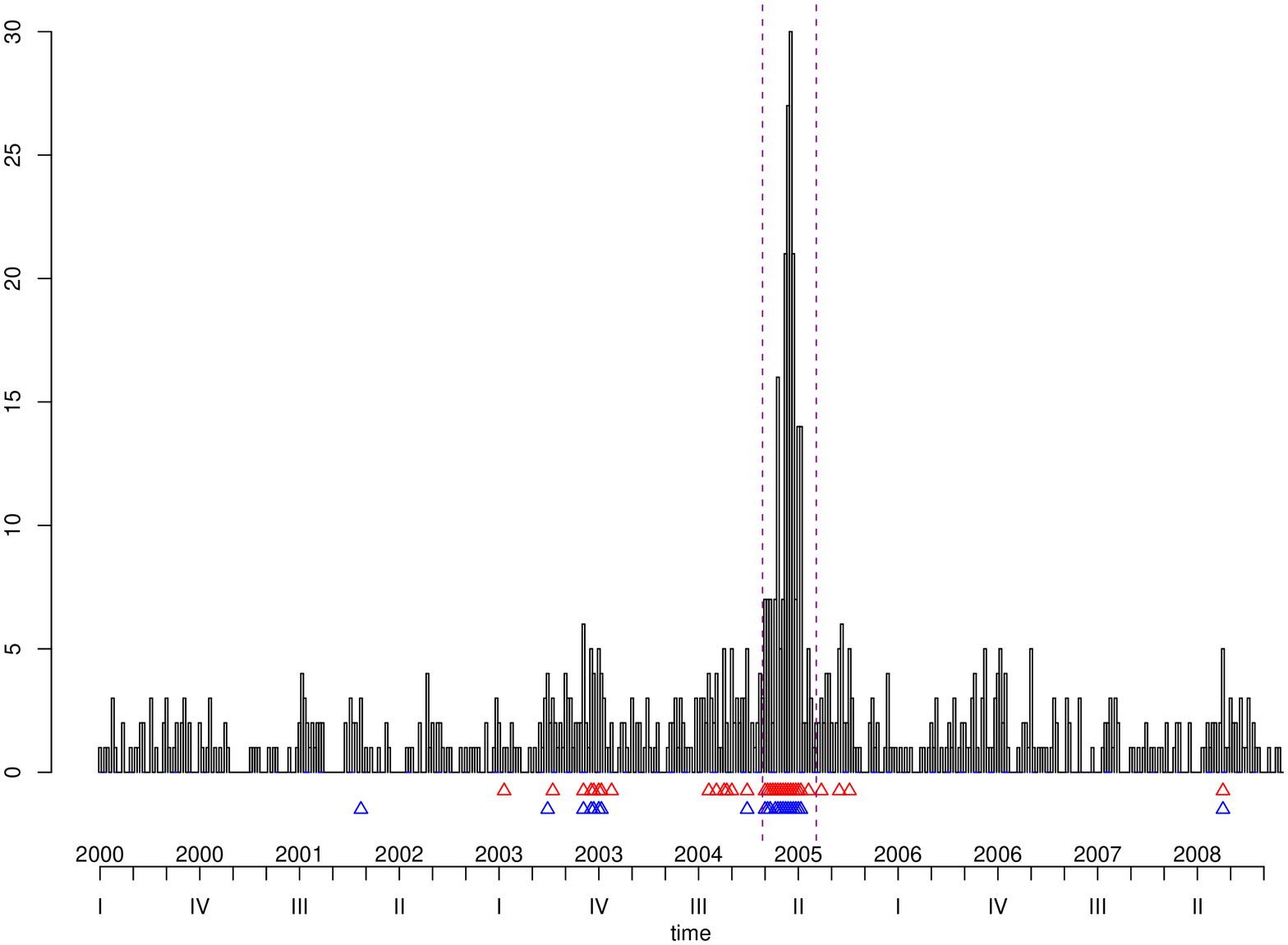}}
\caption{\textit{Salmonella} Agona: Weekly counts from January 1, 2000 to December 31, 2008. Roman numerals refer to the quarters of the years. 
Alarms generated by the EVT method are represented by triangles on the first line. Alarms generated by the Farrington method are represented by triangles 
on the second line. The documented outbreak is delimited by the two dashed lines.}\label{fig:Figure6}
\end{center}
\end{figure}
\noindent In Figure 6, alarms for the serotype Agona are distributed from 2000 to 2008. There is a concordance between the two methods during three periods. The first period, corresponding to 5 weeks in August and September 2003, was not documented as an outbreak. The second concordance period, corresponds to 15 consecutive weeks 
from the last week in December 2004 to week 15 in April 2005. This second period is more interesting as it corresponds to the beginning of a large outbreak of 
infections in infants linked to the consumption of powdered infant formula (see \cite{Brouard2007}). The outbreak period, delimited by the two dashed lines, 
took place in two stages: the first stage from week 53 in December 2004 to week 10 in March 2005 and the second from week 11 to week 21. 
A total of 47 cases less than 12 months age were identified during the first stage and 94 cases less than 12 months age were identified during the second stage. 
The third period corresponds to the week 29 in July 2008. It included five cases, two of them coming from a foodborne disease outbreak involving piglet consumption, 
and the three others being probably sporadic cases. 

\noindent The EVT was implemented using R version 2.9 \cite{R2009}: see \cite{Anis08}. The function called \textit{algo.farrington}, implemented in the R-Package 'surveillance' (\cite{Hoehle2007}) was used to apply the Farrington method.

\section{Discussion}

We believe that the EVT method meets a number of requirements, listed by Farrington {\it et al.} (see \cite{Farrington04}), for the outbreak detection algorithms implemented in surveillance systems. Indeed, this method is able to monitor a large number of time series which became an absolute necessity in modern computerized surveillance systems. 
It can deal with a wide range of events as it is the case for the \textit{Salmonella} infections with the routinely analyses of several hundred serotypes per week. 
It can handle times series with great numbers of cases (such as \textit{Salmonella} Enteritidis) or small numbers of cases (such as \textit{Salmonella} Manhattan). 
Seasonality is taken into account by comparing counts over the same periods of time. Other methods propose a direct way to treat the past aberrations, for instance by associating low weights to the weeks coinciding with past outbreaks. There is no such a need when using the EVT method since the return period is not a constant but depends on each observed count; alarms can then be generated even if past outbreaks exist. It is particularly the case with low counts for which the return period is small and does not include the past outbreaks. 
Finally, the method is implemented in a function using the R language, allowing to run it in an automated procedure with minimal user intervention.\\
\noindent Although the model developed by Farrington {\it et al.} (see \cite{Farrington96}) became a standard reference method, routinely used in France since many years and incorporated in several surveillance systems: human Salmonella, non human Salmonella (see \cite{Baroukh2008}), legionella (see \cite{Grandesso2009}) or in syndromic surveillance systems (see \cite{Fouillet2008}), the EVT method seems also to be a valuable and interesting tool for the recognition of time clusters. It could be integrated in the family of outbreak detection algorithms used by the public health surveillance agencies since developing effective computer-assisted outbreak detection systems still remains a necessity to ensure timely public health intervention.\\~
Another possible way to proceed would be to transform the sample of discrete random variables into a continuous one in order to use standard EVT tools, instead of quantile's bounds. It has been empirically studied when smoothing the data via a kernel transformation and provided promising results (see \cite{Anis08}); it will be the subject of a future work.\\     
Finally, we also plan to investigate the extension of such EVT methods for time or spatially dependent data.

\section*{Acknowledgements}
The authors would like to thank Dr Francois-Xavier Weill, head of the National Reference Center for \textit{Salmonella} in France for providing the datasets and Anis Borchani for the implementation of the method using the R language. \\
{\textit{Conflict of Interest:} None declared.}\\

\end{document}